\documentclass[manuscript,screen]{acmart}
\usepackage{tikz}
\usepackage{natbib}
\usepackage{xcolor}
\usepackage{xargs}
\usepackage{graphbox}
\usepackage{fontawesome5}
\usepackage{listings}
\usepackage{algorithmic}
\usepackage{tcolorbox}
\usepackage{enumitem}
\usepackage{subfigure}
\usepackage{threeparttable} 
\usepackage{booktabs} 

\usetikzlibrary{trees}
\tikzstyle{every node}=[draw=black,thick, anchor=west, minimum height=2.5em, rounded corners=.1cm, font=\sffamily]

\usepackage{graphicx}
\usepackage{subcaption}
\usepackage{array}
\newcolumntype{C}[1]{>{\centering\let\newline\\\arraybackslash\hspace{0pt}}m{#1}}

\definecolor{yesColor}{rgb}{0.565, 0.773, 0.482}
\definecolor{noColor}{rgb}{0.796, 0.447, 0.416}
\definecolor{maybeColor}{rgb}{0.98, 0.73, 0.01}

\newcommand{\YES}{\textcolor{yesColor}{\faCheck}}

\newcommand{\NO}{\textcolor{noColor}{\faTimes}}

\AtBeginDocument{%
  \providecommand\BibTeX{{%
    \normalfont B\kern-0.5em{\scshape i\kern-0.25em b}\kern-0.8em\TeX}}}

\setcopyright{acmlicensed}
\copyrightyear{2024}
\acmYear{2024}
\acmDOI{XXXXXXX.XXXXXXX}

\acmConference[CSUR]{}{June 03--05,
  2024}{Woodstock, NY}
\acmISBN{978-1-4503-XXXX-X/18/06}

\begin{document}

\title{Parallel I/O Characterization and Optimization on Large-Scale HPC Systems: A 360-Degree Survey}


\author{Hammad Ather}
\orcid{0000-0002-0023-8145}
\email{hather@uoregon.edu}
\affiliation{%
    \institution{University of Oregon}
    \city{Eugene}
    \state{Oregon}
    \country{USA}
}

\author{Jean Luca Bez}
\email{jlbez@lbl.gov}
\orcid{0000-0002-3915-1135}
\affiliation{%
    \institution{Lawrence Berkeley National Laboratory}
    \city{Berkeley}
    \state{California}
    \country{USA}
}

\author{Chen Wang}
\email{wang116@llnl.gov}
\orcid{0000-0001-9297-0415}
\affiliation{%
    \institution{Lawrence Livermore National Laboratory}
    \city{Livermore}
    \state{California}
    \country{USA}
}

\author{Hank Childs}
\email{hank@uoregon.edu}
\orcid{0000-0001-5816-1892}
\affiliation{%
    \institution{University of Oregon}
    \city{Eugene}
    \state{Oregon}
    \country{USA}
}

\author{Allen D. Malony}
\email{malony@cs.uoregon.edu}
\orcid{0000-0002-9598-7201}
\affiliation{%
    \institution{University of Oregon}
    \city{Eugene}
    \state{Oregon}
    \country{USA}
}

\author{Suren Byna}
\orcid{0000-0003-3048-3448}
\email{byna.1@osu.edu}
\affiliation{%
    \institution{The Ohio State University}
    \city{Columbus}
    \state{Ohio}
    \country{USA}
}

\renewcommand{\shortauthors}{}

\begin{abstract}

Driven by artificial intelligence, data science, and high-resolution simulations,
I/O workloads and hardware on high-performance computing (HPC) systems have become increasingly complex.
This complexity can lead to 
large I/O overheads and overall performance degradation.
%
%
These inefficiencies are often mitigated using tools and techniques for
characterizing, analyzing, and optimizing the I/O behavior of HPC applications.
That said, the myriad number of tools and techniques available makes it challenging to navigate to the best approach.
In response, this paper surveys $131$ papers from the ACM Digital Library, IEEE Xplore, and other reputable journals to provide a comprehensive analysis, synthesized in the form of a taxonomy, of the current landscape of parallel I/O characterization, analysis, and optimization of large-scale HPC systems.   
We anticipate that this taxonomy will serve as a valuable resource for enhancing I/O performance of HPC applications.

\end{abstract}




    \maketitle

\section{Introduction}
\label{sec:introduction}

High-Performance Computing (HPC) is a critical technology for solving many complex problems.
By leveraging powerful supercomputers composed of state-of-the-art hardware and compute resources, HPC is used to run critical workloads and scientific applications that perform massive computations at a rapid speed. 
Today's HPC systems are capable of large-scale simulations, data processing, and modeling, making them essential for advancing scientific research in domains such as climate modeling, astrophysics, and biology. These systems also have strong applications in defense and national security such as surveillance, cryptography, and cybersecurity. By delivering unparalleled computational power, HPC sets the foundation for driving innovation and scientific discovery.   

Despite the immense power of the current HPC systems, fully unlocking their peak performance remains a key challenge. 
Inefficient usage of an HPC system leads to reduced throughput, leading to a lower return on investment for an expensive resource.
One of the primary causes of inefficiency in HPC systems lies in the complexity of achieving
harmony and coordination between thousands of interconnected
nodes each with different components:
computation, memory, I/O, and networking. 
For an HPC system to operate efficiently, all these components must function harmoniously. A bottleneck in any of these components can severely hamper the performance of the system.
Among these components, the I/O subsystem is often a critical bottleneck,
whether from 
unbalanced I/O workload among processes~\cite{8752753}, 
a large number of concurrent I/O requests~\cite{8752753}, 
redundant/overlapping I/O accesses~\cite{darshan_modular}, 
I/O resource contention at the parallel file system~\cite{Yildiz2016OnTR}, 
or other factors.

To effectively address I/O bottlenecks, it is essential to have a deep understanding of the different methodologies available to analyze and optimize the I/O performance of an application. Evaluating and optimizing parallel I/O is challenging because of the vast amount of performance analysis and optimization tools available. This can lead to the end users facing a dilemma in choosing the most appropriate tool that serves their I/O optimization needs. Moreover, the process of evaluating and optimizing I/O performance is not always straightforward, as it often involves multiple, iterative steps. For example, users may need to select workload generation techniques—such as benchmarks and proxy applications—to accurately replicate the I/O access patterns of their original applications before evaluating I/O performance. With so many potential approaches, identifying the most effective solution for a given situation can be daunting. To simplify this process, a comprehensive resource that synthesizes the different tools, techniques, and their limitations would be highly beneficial to the HPC community. Such a resource would streamline the process of parallel I/O evaluation and tuning, providing clarity and guidance to the end users. By systematically comparing the various tools and methodologies to evaluate and optimize parallel I/O, and consolidating their trade-offs, this resource would significantly help the end users make more efficient and informed decisions for parallel I/O characterization and optimization.



This paper presents such a resource by surveying the current state of the art on parallel I/O evaluation and optimization on large-scale HPC systems. It examines $131$ research papers on parallel I/O workload generation, characterization, and optimization,  synthesizing the findings into a taxonomy for evaluating and optimizing the I/O performance of HPC applications. The paper sifts through the literature available in ACM Digital Library, IEEE Xplore, Springer, Science Direct, and other reputable journals in computer science, as well as some web articles. Overall, the paper has surveyed 37 papers from IEEE Xplore, 22 papers from ACM Digital Library, 8 papers from Springer, 4 from Science Direct, and around 40 papers from other reputable journals worldwide. It also surveys 20 different websites and articles. It provides in-depth details of the different layers of the HPC I/O stack and examines how I/O access patterns, identified in previous research \cite{iosurvey}, affect performance as data requests move through the stack. After looking at the HPC I/O stack and the different access patterns, the paper classifies the various stages of parallel I/O evaluation and tuning, including workload generation, profiling, and monitoring. It also explores various analysis and optimization methods-including statistical analysis, machine learning based analysis, and replay-based modeling, and discusses targeted optimization techniques for the different layers of the I/O stack. Through this survey, this paper addresses the following key questions: 

\begin{itemize}
\item What are the different layers of the HPC I/O stack and how do they interact with each other?
\item What profiling and tracing tools are available to characterize application I/O, and what are their limitations?
\item Which tuning techniques are applied by HPC researchers for optimizing the I/O behavior of the application?
\end{itemize}

The rest of the paper is organized as follows. Section \ref{sec:hpc-stack} looks at the different layers of the HPC I/O stack. Section \ref{sec:taxonomy} introduces the node-link hierarchical tree diagram to describe a parallel I/O evaluation and optimization taxonomy. Section \ref{sec:parallel-io-evaluation} and \ref{sec:parallel-io-optimization} discuss the different stages of parallel I/O evaluation and optimization. Section \ref{sec:gaps} discusses limitations and gaps in the current workload generation, I/O characterization, and optimization techniques. Lastly, Section \ref{sec:conclusion} presents the conclusion of this survey. 

\section{HPC I/O Stack}
\label{sec:hpc-stack}
The complex I/O workloads from serial and parallel applications running on large-scale HPC systems are supported by the HPC I/O stack \cite{iosurvey}. The I/O stack is complex and has multiple layers, with each layer having a multitude of tuning parameters that can used to improve the performance of the application \cite{ioautotuning, dxt-explorer}. Applications must traverse multiple layers to reach the storage hardware: high-level I/O libraries, parallel I/O middleware, low-level I/O libraries, I/O forwarding layer, and the parallel file system (PFS).

\subsection{High-level I/O Libraries}

High-level I/O libraries provide abstractions for data modeling and management, allowing portability and high application performance. Some of the commonly used high-level I/O libraries include HDF5 \cite{hdf5}, ADIOS \cite{ADIOS}, NetCDF \cite{NetCDF-4}, and PnetCDF \cite{PnetCDF}. The HDF5 technology suite comprises a data model, a library, and a file format to store and manage data \cite{HDF5suite}. NetCDF (Network Common Data Form) supports creating, accessing, and sharing array-oriented scientific data using software libraries and machine-independent data formats \cite{netcdf}. PnetCDF uses parallel I/O techniques to provide a high-performance and efficient interface to access netCDF files. The Adaptable IO System (ADIOS) provides an efficient and flexible solution for researchers and scientists to describe the data in their code as an external-to-the-code XML file for processing outside the running simulation. Apart from these high-level I/O libraries, there are also some domain-specific I/O libraries as well such as FITS \cite{FITS, FITSarticle} and ROOT \cite{ROOT}, which are used in astronomy and High-Energy Physics (HEP) respectively. 

\subsection{Parallel I/O Middleware and Low-Level I/O Libraries}

The parallel I/O middleware and low-level I/O libraries include the MPI-IO (Message Passing Interface I/O) \cite{MPI-IO}, POSIX I/O (Portable Operating System Interface I/O) \cite{POSIX}, and STDIO (Standard Input and Output) interfaces. In MPI-IO, a file is an ordered collection of typed data items. Using these typed data items, MPI-IO allows the user to define data models for their applications. It also provides two types of I/O calls, which are independent I/O and collective I/O. 
POSIX I/O views a file as a sequence of bytes. Through this interface, contiguous regions of bytes can be transferred between the file and memory. Noncontiguous regions of bytes can be transferred from memory to a file by giving complete low-level control of the I/O operations. The major drawback of POSIX I/O in the context of HPC is its strong consistency model \cite{10.1145/3431379.3460637, 10504997} and little support for parallel I/O. It is also not flexible regarding file metadata as it prescribes a specific set of metadata that a file must possess \cite{Prickett_2018}. STDIO, in contrast, provides abstractions to deal with the stream of input and output bytes. It relies on the user to create an input/output stream, seek the position in the file from where to read or write, and then read/write bytes in sequence from/to the stream.

\subsection{I/O Forwarding Layer}

The I/O forwarding layer technique was initially introduced for the Blue Gene/L system \cite{blue-gene}. The purpose of this layer is to reduce the number of clients accessing a PFS concurrently by introducing a transparent intermediary layer between the compute nodes and the data servers. It achieves this goal by using I/O nodes. Instead of the applications directly accessing the PFS, requests are first received by these I/O nodes, which forward these requests to the PFS in a manageable way. This technique allows a smooth flow of I/O requests, providing a layer between the application and file system. It enables application-wise and system-wide optimization techniques such as aggregation, request scheduling, and compression to have a broader view, considering the accesses from multiple compute nodes.

\subsection{Parallel File System}

Large-scale HPC systems rely heavily on PFS to provide an infrastructure for a unified global namespace and persistent shared storage, to enable fast and efficient read and write operations on files across distributed storage servers. A PFS mainly has two types of servers: the data server and the metadata server. Metadata servers are responsible for handling the file metadata, which is information related to the size, permissions, and location of (parts of) the file on the data servers. Before any client can access the data in a file, they must obtain the layout information of the file from the metadata. On the data server, the files are distributed using data striping \cite{striping}. In the technique, the file is divided into chunks called stripes, distributed to the servers in a round-robin approach. A PFS can retrieve stripes from different servers in parallel, increasing throughput. For instance, PVFS \cite{PVFS}, Lustre \cite{2003LustreA, osti_1824954}, and Spectrum Scale (previously GPFS \cite{GPFS}) have default stripe sizes between 64KB and 1MB. However, some PFS systems use locking on the server to maintain consistency with concurrent access, which could impact scalability. Lustre, PVFS, IMB Spectrum Scale, and the Panasas file system \cite{panasas} are some of the commonly used PFS. 

\subsection{Storage Hardware}

Storage devices are the last layer of the I/O stack. There are a variety of storage hardware used in a supercomputer. Hard Disk Drives (HDDs) \cite{hdds} are a popular storage medium that has been in use for many years now. HDDs perform best when the underlying data needs to be accessed sequentially instead of randomly, as that reduces the overhead of the seek operation \cite{Patterson_2013}. A recent flash-based alternative to hard disks is Solid-State Drives (SSDs). SSDs have higher bandwidth and less overhead for sequential access than HDDs. NVMe (Non-Volatile Memory Express) is another high-speed storage interface and protocol designed to optimize the performance of SSDs by reducing latency and maximizing data transfer rate. 

\begin{tcolorbox}[size=title, top=5px, bottom=5px, boxrule=0.1mm, arc=0.5mm, colback=white, colframe=black, fonttitle=\sffamily\bfseries, title=Summary \#1]
   The I/O stack deployed on large-scale HPC systems is complex. It comprises multiple layers with numerous tuning parameters aimed at improving application performance and supporting the diverse requirements of HPC systems. High-level I/O libraries like HDF5 \cite{hdf5}, NetCDF \cite{NetCDF-4}, PnetCDF \cite{PnetCDF}, and ADIOS \cite{ADIOS} provide data modeling and management abstractions, enhancing portability and performance. Then, parallel middleware, which includes MPI-IO \cite{MPI-IO}, provides parallel I/O support, and low-level I/O libraries, such as POSIX \cite {POSIX} and STDIO, provide interfaces for file handling. The I/O forwarding layer is introduced to optimize access to PFS and acts as an intermediary between compute nodes and data servers. PFS provides persistent shared storage across distributed servers, divided into data and metadata servers. Storage hardware provides mediums such as HDDs and SSDs and interfaces such as NVMe.
\end{tcolorbox}


\section{Parallel I/O Evaluation and Optimization Taxonomy}
\label{sec:taxonomy}
\begin{figure*}[ht]
    \centering
    \includegraphics[width=0.92\columnwidth]{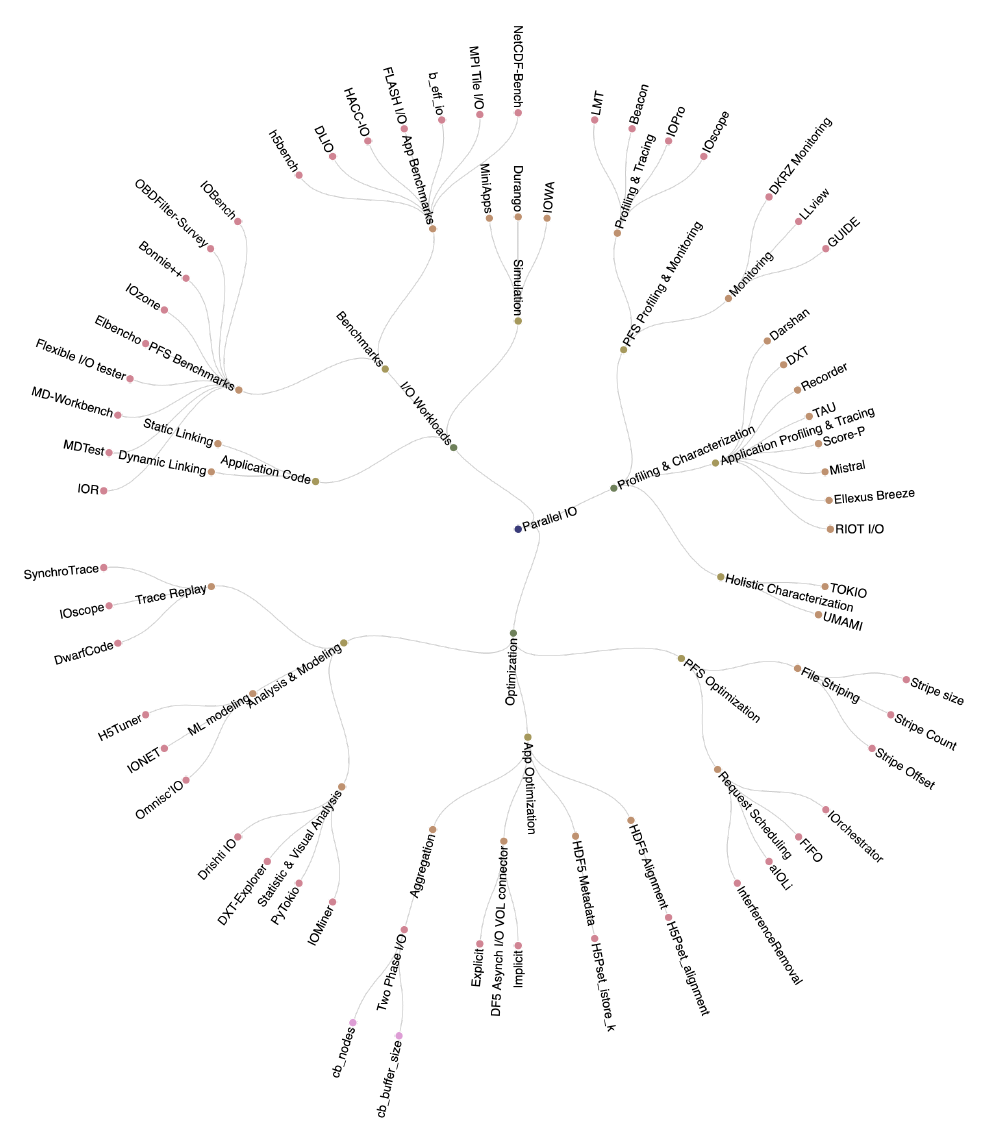}
    \caption{The node-link hierarchical tree diagram to describe a taxonomy for parallel I/O evaluation and optimization of large-scale HPC systems. The taxonomy shows different phases of parallel I/O evaluation and presents a comprehensive list of workload generation techniques, profiling, and tracing tools for I/O characterization. It also presents various analysis and optimization methods, including statistical analysis, predictive analysis, and replay-based modeling, and discusses targeted optimization techniques for the different layers of the I/O stack.}
    \Description{Taxonomy}
    \label{fig:parallel-io-eval}
\end{figure*}

Parallel I/O evaluation and optimization is a complex task involving multiple steps. This survey presents a node-link hierarchical tree diagram to describe a taxonomy for parallel I/O evaluation and optimization of large-scale HPC systems. This taxonomy, shown in Fig. \ref{fig:parallel-io-eval}, captures key aspects such as workload generation, data monitoring and collection, and performance analysis and optimization. Workload generation involves generating, capturing, or identifying the I/O workload. The I/O workload can be an existing application for which the I/O behavior needs to be investigated. However, if the application or its code is unavailable, there are methods for generating I/O workloads, which include benchmarks, workload replication, and simulation frameworks. Data monitoring and collection includes profiling, tracing, and monitoring tools which are used to understand where the performance issues are. Lastly, performance analysis and optimization focus on using tuning methods to get rid of or at least reduce the impacts of I/O bottlenecks and improve both the application's and overall system's performance. This multi-step process (from generating I/O workloads to applying optimizations) makes parallel I/O evaluation and optimization a complex iterative workflow, which, if not understood properly, can leave the HPC system with I/O issues and bottlenecks. 
 
The paper discusses parallel I/O performance evaluation and optimization over two complementary facets: \textbf{System} and \textbf{Application}. For system-level performance evaluation and optimization, this survey examines various tools to benchmark parallel file and storage system performance. It also looks at various profiling, tracing, and monitoring systems that can be utilized to assess system-wide I/O performance. It also covers tuning techniques aimed at optimizing I/O performance at the system level. For application-level evaluation and optimization, the paper reviews various benchmarks that can emulate different access patterns and application behavior, generating workloads that mirror real-world scenarios. It also looks at various profiling, tracing, and monitoring systems that can be utilized to assess application performance. Finally, the paper outlines optimization strategies to enhance application-level I/O efficiency.

\section{Evaluating Parallel I/O}
\label{sec:parallel-io-evaluation}
\subsection{I/O Workloads}

\subsubsection{Application Code}
The most accurate workload that can be used for characterization is the application code. Many profiling tools dynamically link with the application to instrument it and collect performance data. However, if the profiling tool cannot dynamically instrument the application, it might statically link to the application for instrumentation. In many cases, profiling and tracing application code to understand I/O behavior is not feasible, especially for large-scale, complex scientific applications. Doing so can introduce significant overhead and require managing a complex set of dependencies. In such cases, end-users and developers often rely on other workload generation methods, such as benchmarks, simulation frameworks, and proxy applications. 

\subsubsection{Benchmarks}
Benchmarks simulate the workflow and access patterns of the original application, enabling performance evaluation and tuning without the overhead of running the full application. Benchmarks often abstract computation and communication but are still able to represent communication incurred from I/O (e.g., when collective I/O is used in MPI-IO). In this section, we analyze a range of system and application benchmarks to uncover their key characteristics. 

\paragraph{Parallel File System and Storage System Benchmarks}
PFS and storage systems play a critical role in determining the I/O performance of an HPC system \cite{10.1016/j.jpdc.2023.104744, 7405579}, making it crucial to understand and optimize the performance of these sub-systems. Many system-focused benchmarks are available that exercise different aspects of system performance, such as file system, metadata, storage, and disk performance. We examine these benchmarks and summarize our findings in Table \ref{tab:system_benchmarks}.

IOR \cite{ior_documentation} is a synthetic parallel I/O benchmark based on MPI that uses different interfaces and access patterns to test the performance of storage systems. The benchmark provides over 50 options to generate various workloads. However, the main focus is to measure the sequential read/write performance under different parameters such as file size, I/O transfer size, and concurrency. It also provides options for using a shared or one file per process. Additionally, it supports both the conventional POSIX interface and the parallel MPI-IO interface, and high-level libraries such as HDF5. Various research studies \cite{IOR, Madireddy17b} have used IOR to understand the performance of the parallel storage system. \citet{IOR} characterizes and predicts the I/O performance of HPC applications using the IOR benchmark and presents a procedure for selecting IOR parameters to generate I/O patterns that match exactly with the selected applications. These I/O patterns can then be used to predict the I/O performance of the original application. 


Metadata performance is a critical measurement in achieving the full capabilities of a PFS \cite{MDTest, 10.1145/3149393.3149401}. MDTest \cite{MDTest} is a MPI-based I/O benchmarking tool, which is now a part of IOR, and is used to test the metadata performance of the storage system. MDTest demonstrates the peak rate at which the file system can perform operations such as open, close, stat, mkdir, create, and unlink. MD-Workbench benchmark \cite{MD-Workbench} is another MPI-parallel benchmark to measure metadata performance. This benchmark aims to mimic the concurrent access to typically small objects and different user activities on a file system. In contrast to MDTest, MD-Workbench can simulate access patterns that are challenging to cache and optimize for existing file systems. Consequently, its performance is significantly lower than MDTest, achieving around 10k IOPS as compared to 100k IOPS with MDTest.  

Flexible I/O tester (fio) \cite{fio} is a useful benchmarking tool for testing the performance of Linux kernel I/O interfaces. It provides a broad range of plugins and libraries to test the single-node performance for storage devices. Elbencho \cite{elbencho} is a new storage benchmark that can simulate a large variety of access patterns and can even include GPUs in the data accesses. It aims to provide a modern and flexible storage benchmarking tool by combining the best features of IOR and fio while providing a live text UI to monitor real time I/O performance. Elbencho can simulate a broad range of test cases, such as testing for a lot of small files with varying sizes, access latency, IOPS for random access to shared files, and others. 

IOzone \cite{IBM} is another benchmark to analyze file system performance. The benchmark can not only generate, but also measure different file operations such as read, write, spread, read, read backward, read strided, fwrite, and random read. IOzone can run on multiple operating systems and be easily ported to other machines. Bonnie++ \cite{bonniepp} is a popular tool used for benchmarking file system performance, specifically focusing on measuring I/O performance on Unix-like systems. The benchmark can test metadata operations like file creation and deletion as well as measure the data throughput for both random and sequential read/write operations.

OBDFilter-Survey \cite{OBDFilter} is a Lustre benchmarking tool to measure the performance of Object Storage Targets (OSTs) attached to an  Object Storage Service (OSS). It simulates a Lustre client's I/O patterns by using a script that generates sequential I/O from various objects and threads. OST storage performance can be measured without any intervening network by running OBDFilter-Survey directly on an OSS node. OST performance, including network overhead, can also be measured by running OBDFilter-Survey remotely on a Lustre client. OBDFilter-Survey does not work well when the system scales beyond a specific number of OSTs, as it only measures an individual storage subsystem's I/O performance \cite{OBDFilter}. 

IOBench \cite{IOBench} is a system I/O performance benchmark that supports several I/O engines such as blocked IO, Linux asynchronous IO and POSIX asynchronous I/O (AIO). The benchmark measures the sequential write, random write, backward write, sequential read, random read, and backward read performance to benchmark storage systems or investigate issues with storage protocols or local devices. This tool can be easily scaled to many storage devices and CPUs.

\begin{table*}
  \begin{tabular}{lccccc}
    \textbf{Benchmark} & \textbf{Data} & \textbf{Metadata} & \textbf{GPU Support} & \textbf{File System} & \textbf{Storage System} \\
    \toprule
     IOR \cite{IOR} & \YES & \NO & \NO & \YES & \YES\\
     Mdtest \cite{MDTest} & \NO & \YES & \NO & \YES & \NO \\
     MD-Workbench \cite{MD-Workbench} & \NO & \YES & \NO & \YES & \NO \\
     Fio \cite{fio} & \YES & \NO & \NO & \NO & \YES \\
     Elbencho \cite{elbencho} & \YES & \NO & \YES & \YES & \YES \\
     IOzone \cite{IBM} & \YES & \NO & \NO & \YES & \NO\\
     Bonnie++ \cite{bonniepp} &  \YES & \NO & \NO & \YES & \YES\\
     OBDFilter-Survey \cite{OBDFilter} & \YES & \NO & \NO & \YES & \NO\\
     IOBench \cite{IOBench} & \YES & \NO & \NO & \NO & \YES \\
    \bottomrule
  \end{tabular}
  \caption{List of I/O benchmarks to exercise parallel file systems and storage systems} 
\label{tab:system_benchmarks}
\end{table*}

\paragraph{Application Benchmarks}

This section looks at various application benchmarks and determines their key characteristics. It focuses on how these benchmarks represent different I/O behaviors, including their relevance to scientific applications and the types of workloads they simulate. The analysis also covers the layers of the I/O stack each benchmark evaluates. The findings, summarized in Table \ref{tab:app_benchmarks}, provide a comparison of the benchmarks in terms of their coverage, type, and I/O access patterns.

h5bench \cite{h5bench} provides a versatile suite of HDF5 I/O kernels or benchmarks (e.g., OpenPMD \cite{openpmd}, AMReX \cite{10.1177/10943420211022811}, E3SM \cite{e3sm}, MACSio \cite{MACSio}) that represent different applications using the HDF5 and parallel I/O libraries, measuring various aspects of I/O performance such as the I/O overhead and observed I/O rate. Among the benchmarks provided by h5bench, the read and write benchmarks are the most basic ones to get started with and represent the patterns observed in VPIC-IO \cite{vpic} and BDCATS-IO \cite{bdcats}. The user can exercise various I/O patterns in these benchmarks through configurable options.

Deep Learning I/O benchmark (DLIO) \cite{DLIO} aims to emulate the behavior of scientific deep learning applications. It characterizes the I/O behavior of these applications with the goal of data-centric optimizations on modern HPC systems. Delivered as an executable and using a modular design to incorporate many datasets, data formats, and configuration parameters, the benchmark allows the user to configure various I/O patterns, like I/O mode, spatiality, and file approach. 

HACC-IO \cite{HACC-IO} evaluates the I/O performance of the Hardware Accelerated Cosmology Code (HACC) \cite{HACC-IO} simulation. HACC is a high-performance cosmological simulation code, designed for extreme-scale computing, that models structure formation in collision-less fluids influenced by gravity within an expanding universe using N-body techniques. The HACC I/O benchmark aims to capture the different I/O patterns of the HACC simulation code, such as the different I/O interfaces like POSIX I/O and MPI-IO (collective and independent). The benchmark can also replicate other access patterns, such as writing out a single shared file, file per process, and a file per group of processes.

The FLASH I/O benchmark evaluates the performance of the FLASH parallel HDF5 output \cite{Flash}. Using the parallel HDF5 library and recreating the data structures in FLASH, the benchmark produces three output files: a checkpoint file, a plotfile with centered data, and a plotfile with corner data. FLASH I/O provides identical I/O routines used by FLASH and aims to tune the I/O performance in a controlled environment by focusing on the write performance. 

The effective I/O bandwidth benchmark (b\_eff\_io) \cite{beffio} evaluates the average I/O bandwidth achievable by parallel MPI-I/O applications using different access patterns, access methods, and buffer lengths. It examines various access patterns on one file per application such as first write, rewrite, strided, and segmented collective patterns. It also examines non-collective accesses to one file per process. Lastly, it can also vary the number of parallel accessing processes to evaluate I/O performance.

MPI Tile I/O \cite{tile_i/o} benchmark tests the performance of an underlying MPI-IO and file system implementation under a noncontiguous access workload. It divides a data file into a dense two-dimensional set of tiles, and the user can define the number of tiles in $x$ and $y$ dimensions as well as the number of elements in each dimension. The benchmark provides parameters to control the behavior of the application, such as collective I/O and the number of MPI I/O aggregators that can be used per node. 

NetCDF Performance Benchmark Tool (NetCDF-Bench) \cite{NETCDF-Bench} aims to measure NetCDF performance on a range of devices such as notebooks to large HPC systems. By mimicking the I/O behavior of scientific climate applications, it captures the performance of each node/process and provides a human-readable summary. NetCDF-Bench is a parallel benchmark supporting independent I/O, collective I/O, and chunked I/O modes.

\begin{table*}
\begin{tabular}{lccc|cc|cccc}
    \multicolumn{1}{c}{} & \multicolumn{3}{c}{\textbf{Layer}} & \multicolumn{2}{c}{\textbf{Type}} & \multicolumn{4}{c}{\textbf{Access Pattern}} \\
     \textbf{Benchmark} & \rotatebox{90}{\textbf{\small Low Level}} & \rotatebox{90}{\textbf{\small Middleware}} & \rotatebox{90}{\textbf{\small High Level}} & \rotatebox{90}{\textbf{\small Synthetic}} &  \rotatebox{90}{\textbf{\small Application}} & \rotatebox{90}{\textbf{\small I/O Mode}} &  \rotatebox{90}{\textbf{\small Operation}} & \rotatebox{90}{\textbf{\small Spatiality}} & \rotatebox{90}{\textbf{\small File Approach}}\\
    \toprule
     h5bench \cite{h5bench} & \YES & \YES & \YES & \YES & \YES & \YES & \YES & \YES & \NO \\
     DLIO \cite{DLIO} & \NO & \NO & \YES & \YES & \NO & \YES & \NO & \YES & \YES \\
     HACC-IO \cite{HACC-IO} & \YES & \YES & \YES & \YES & \NO & \YES & \NO & \NO & \YES \\
     FLASH-IO \cite{Flash} & \NO & \NO & \YES & \YES & \NO & \NO & \YES & \NO & \NO \\
     b-eff-io \cite{beffio} & \YES & \YES & \NO & \YES & \NO & \YES & \YES & \YES & \YES \\
     MPI Tile I/O \cite{tile_i/o} & \NO & \YES & \NO & \YES & \NO & \YES & \YES & \NO & \NO \\
    NetCDF-Bench \cite{NETCDF-Bench} & \NO & \NO & \YES & \YES & \NO & \YES & \YES & \NO & \NO \\
    Parabench \cite{parabench} & \YES & \YES & \NO & \YES & \NO  & - & - & - & - \\
    \bottomrule
  \end{tabular}
  \caption{List of HPC I/O application benchmarks along with some of their characteristics} 
\label{tab:app_benchmarks}
\end{table*}

\subsubsection{Workload Replication and Simulation Frameworks}

Proxy applications are small and simplified codes of large and complex production applications that encapsulate the important features of these applications without forcing the user to assimilate large and complex code bases. \citet{proxy_1} presents proxy applications called MiniApps, which are based on large-scale application code at the Oak Ridge National Lab (ONRL). Encapsulating all the important features and details of these large-scale applications, these MiniApps run on production systems at ORNL. \citet{Dickson2017EnablingPI} is another work that replicates five different I/O workloads using MACSio \cite{MACSio} proxy application and Darshan \cite{darshan}. 

\citet{10.1145/3064911.3064923} provides a novel framework, called Durango, to generate scalable workloads from real applications using the performance modeling language Aspen and the HPC CODES framework \cite{codes}. \citet{10.1145/2832087.2832091} presents IOWA, a novel I/O workload abstraction for generating diverse I/O workloads based on the inputs sources. \citet{7836562} extracts the I/O pattern of the application and generates a suitable proxy application.

\begin{tcolorbox}[size=title, top=5px, bottom=5px, boxrule=0.1mm, arc=0.5mm, colback=white, colframe=black, fonttitle=\sffamily\bfseries, title=Summary \#2]
This section explores various methods for generating and capturing the I/O workloads in an HPC system. It begins by discussing the use of application code as the most accurate workload for characterization, though profiling large-scale applications may introduce significant overhead. To address this, benchmarks simulate the I/O behavior of applications, enabling performance evaluation without incurring the overhead of running the full application. A variety of system and application benchmarks, such as IOR \cite{IOR} and h5bench \cite{h5bench}, are discussed in this section and the findings are summarized in Table \ref{tab:system_benchmarks} and \ref{tab:app_benchmarks}. Finally, the section ends the discussion by talking about proxy applications and simulation frameworks like MiniApps \cite{proxy_1}, Durango \cite{10.1145/3064911.3064923}, and IOWA \cite{10.1145/2832087.2832091} which are used to replicate complex workloads and generate diverse I/O patterns for testing, providing valuable insights into performance optimization. 
\end{tcolorbox}

\subsection{I/O Profiling and Characterization}

Once the I/O workloads are generated through the different means discussed in the previous section, profiling, and monitoring are the next steps in I/O performance evaluation. This includes collecting application I/O performance data, such as time spent in I/O operations, read and write throughput, and total execution time. This data is then used to characterize the application's I/O performance and determine tuning strategies. This performance data is also used for various analysis, modeling, and prediction techniques that help in optimizing I/O performance. 

A variety of I/O profiling and monitoring tools are available to help understand the I/O behavior of the application and parallel file and storage systems. Each tool has different characteristics and ways of instrumenting I/O workloads. This section discusses the different I/O profiling, tracing, and monitoring tools available to the HPC community. It divides the tools into two categories: (1) Parallel File and Storage System Profiling Tools and (2) Application Profiling Tools. Lastly, this section also looks at some recent monitoring systems that provide an all-encompassing and cohesive view of the I/O behavior of the application.

\subsubsection{Parallel File and Storage Systems Profiling, Tracing, and Monitoring}

Various profiling, tracing tools, and monitoring tools are available to examine the I/O performance of parallel file and storage systems. This section looks at some of these tools and summarizes the findings in Table \ref{tab:sys_profiling}. 

\paragraph{Profiling and Tracing}

Lustre Monitoring Tool (LMT) \cite{Lustre} is an open-source tool that monitors Lustre activity on HPC systems using the Cerebro monitoring system and provides a ``top''-like display of activity on server-side nodes: Object Storage Server (OSS), Metadata Server (MDS), and Lustre Networking (LNET) routers.  It also provides a MySQL/MariaDB to store historical data pulled from the Cerebro Monitoring System. LMT provides text clients that can display real-time data from Cerebro and also graph historical data stored in the MySQL database. It is also possible to directly mine the MySQL database, e.g., with SQL scripts and GNU plots. LMT provides information such as bytes read/written, CPU load averages, and metadata operation rates.

\begin{table*}
\begin{tabular}{l|ccc|cc}
    \multicolumn{1}{c}{} & \multicolumn{3}{c}{\textbf{Type}} & \multicolumn{2}{c}{\textbf{Stack Layer}} \\
     \textbf{Benchmark} &\textbf{\small Profiling} & \textbf{\small Tracing} & \textbf{\small Monitoring} & \textbf{\small Parallel File System} &  \textbf{\small Storage}\\
       \toprule
    LMT \cite{Lustre} & \NO & \NO & \YES & \YES & \YES \\
    Beacon I/O \cite{Beacon} & \NO & \NO & \YES & \YES & \YES \\
      IOPin \cite{IOPin} & \YES & \NO & \NO & \YES & \NO \\
     IOPro \cite{IOPro} & \YES & \NO & \NO & \YES & \NO \\
     IOscope \cite{10.1007/978-3-030-02465-9_7} & \NO & \YES & \NO & \NO & \YES \\
     DKRZ Monitoring \cite{10.1007/978-3-030-02465-9_4} & \NO & \NO & \YES & \YES & \YES \\
     LLview \cite{LLview} & \NO & \NO & \YES & \YES & \YES \\
     GUIDE \cite{9926255} & \NO & \NO & \YES & \YES & \YES \\
    \bottomrule
  \end{tabular}
  \caption{List of parallel file and storage system profiling, tracing, and monitoring tools and their characteristics}  
\label{tab:sys_profiling}
\end{table*}

There are three main components of the Lustre Monitoring Tool. The first is lmt-server-agent package, which contains Cerebro metric plugins that run on Lustre servers, reading Lustre /proc and /sys values and pushing that data into the Cerebro monitoring network. The lmt-server contains a Cerebro monitor plugin that is responsible for pulling data from the Cerebro monitoring network and (optionally) pushing it into the MySQL/MariaDB database. It also provides interfaces that administer LMT such as ltop to display live data and lmtsh to display historical data from the database. The lmt-gui package provides interfaces like lwatch, which provide a GUI for LMT. However, this package is no longer maintained, making it obsolete. 

Beacon \cite{Beacon} is an end-to-end I/O monitoring and diagnosis tool developed for the TaihuLight supercomputer \cite{TaihuLight}. It monitors and correlates I/O trace and profile data collected on different types of nodes, such as the compute, I/O forwarding, metadata, and storage nodes. At each node, Beacon deploys a daemon to monitor I/O and collect performance data to be later used for aggregation. It uses an aggressive first-pass compression technique on all compute nodes to collect and store per-application I/O trace data efficiently and uses log servers as its major processing and storage units. These servers are hosted on the storage nodes, utilizing hardware resources and parallelism \cite{Beacon}. The log servers adopt a layered architecture to collect I/O performance and quickly absorb monitoring output data using Logstash \cite{Sanjappa2016AnalysisOL} and Redis \cite{redis}. Furthermore, it uses Elasticsearch to analyze data on persistent data storage. Finally, it presents the monitoring results to the user using a beacon server, which provides a web interface for data visualization, per job I/O information summary, and an in-memory cache for recent jobs. 

IOPro \cite{IOPro} is a framework for analyzing and visualizing the performance of parallel I/O HPC applications. What sets IOPro apart from other tools is its ability to instrument the entire parallel I/O stack, providing a comprehensive view of I/O performance across all layers. Instead of manually instrumenting each layer of the stack, it takes the I/O software stack as input and generates an instrumented I/O stack to trace specific I/O functions and individual components across each layer. It has three main components: \textit{instrumentation engine}, \textit{execution engine}, and \textit{data processing engine}. The instrumentation engine is made up of a probe selector and inserter. By automatically inserting these probes into the I/O software stack, it generates an instrumented version of the stack. The execution engine is responsible for building and compiling the instrumented software stack. The data process engine is responsible for collecting all trace log files from each layer of the instrumented
I/O stack. 

IOPin \cite{IOPin} is a dynamic instrumentation tool developed to analyze the performance of parallel I/O. It uses Pin \cite{Pin}, which is a lightweight binary instrumentation tool to intercept the MPI-IO and PFS layers. Because of this, IOPin does not require any recompilation of the source code or the I/O software stack. The basic architecture of IOPin is presented in \citet{IOPin}, which exemplifies how IOPin performs dynamic instrumentation when a collective MPI-IO write call is issued. In that architecture, trace information for the MPI and PFS layers is generated using two Pin profiling processes on the client and server sides. The client-side Pin process contains information such as the MPI call ID, rank, PFS call ID, I/O type (read/write), and latency. The server-side Pin process stores information such as PFS server ID, latency in the server, the number of disk accesses, bytes to be read/written, and disk throughput for the MPI I/O call at runtime.

IOscope \cite{10.1007/978-3-030-02465-9_7} is a flexible I/O tracer for characterizing the I/O patterns of the storage systems' workloads. It relies on the extended Berkeley Packet Filter (eBPF) technology \cite{9968265} to instrument the kernel and communicate the target kernel data towards a userspace process. The tool detects the order in which the file offsets are accessed by the I/O requests during accessing on-disk data. \emph{IOscope} consists of two tools: \emph{IOscope classic} and \emph{IOscope mmap}. The main idea is to trace and filter the workloads' I/O requests to construct the workloads' I/O patterns. 

\paragraph{Monitoring}

The German Climate Computation Centre (DKRZ) developed a monitoring system \cite{10.1007/978-3-030-02465-9_4} to monitor the workload of the Mistral Supercomputer. 
This system is built of open-source components such as Grafana \cite{grafana} and also includes a self-developed data collector. It provides statistics about the login nodes, user jobs, and the workload manager queue. Initially, the monitoring system provides users with an overview of the system's current state, providing information such as the current load on login nodes and the number of nodes in use across Slurm partitions. For each node in the system, the monitoring system collects metrics such as CPU usage, memory usage, luster file system usage, and energy consumption. 

The Julich Supercomputing Centre (JSC) developed LLview \cite{LLview} to provide an interactive graphical tool to monitor jobs running on different workload managers such as Slurm. It has a job reporting module, which provides detailed information about all the individual jobs running on the system. The way LLview works is that it interacts with the different sources in the system to collect performance data and aggregate that data to present to the user in a web portal. Apart from providing a live view of the system, this web portal provides a list of the jobs running on the system in the form of a table, with each row showing the aggregated information for that job, such as the owner, project, start time, end time, etc.  Upon clicking on a job, the user can also see timeline graphs for the key performance metrics. All metrics in job reporting are gathered every minute. 

GUIDE (Grand Unified Information Directory Environment) \cite{9926255} is another framework that has been developed to collect, integrate, and analyze log data from the Oak Ridge Leadership Computing Facility (OLCF). The framework first collects data at each level of the OLCF subsystems. For example, at the storage subsystem level, data is collected for the disk layer, the redundant RAID, controllers, the OSSes, and the Lustre PFS level. In the next step, the data is preprocessed using data cleansing techniques. In the federation step, the data is federated in a scalable repository to make post-processing and visualization of the data easier. Lastly, in the post-processing step, a suite of analytics and post-processing techniques are applied to get insights from the data. 


\subsubsection{Application Profiling, Tracing, and Monitoring}

This section looks at various profiling, tracing, and monitoring tools that collect data for high-level, parallel middleware, and low-level I/O libraries such as HDF5, MPI-IO, and POSIX to help understand application performance. It studies different characteristics of these tools, such as the kind of data they collect, their instrumentation technique, and which layer of the stack they instrument, summarizing these findings in Table \ref{tab:app_profiling}. 

Darshan \cite{darshan, darshan_modular} is a commonly used application I/O characterization tool deployed on many large-scale HPC systems. It collects fine-grain trace data for POSIX and MPI-IO layers (parallel and low-level I/O layer) and collects basic data for high-level I/O libraries such as HDF5 and PNetCDF. Implemented as a user-space library, Darshan does not require any source code modification to instrument the application. Instead, it can be added to the application either statically or dynamically. Dynamic executables use the LD\_PRELOAD environment variable to inject Darshan instrumentation in the application, whereas static executables link Darshan to the application during the linking phase. Instead of working as a typical tracing tool that logs every I/O operation, a bounded amount of data for each file is collected by Darshan, allowing compact storage of data. This data contains metrics such as I/O operations count, I/O access sizes, timers, and other statistical data relevant to the I/O performance of the application. 



During the application execution, Darshan instruments different layers of the I/O stack for each process, generating data records characterizing the application’s I/O workload and writing that data to the process's file while deferring all communication and I/O operations until the application terminates. When the application terminates, as part of the darshan\_shutdown routine, it performs data aggregation and compression, which comprises aggregating and compressing the data of all the file-per-process into a single file. Darshan does this process in parallel and uses MPI-IO collective writes to make this step faster. This ability to delay communication and I/O until application shutdown, combined with the bounded data collection, makes Darshan lightweight and portable, allowing seamless deployed on large-scale HPC systems such as the Argonne Leadership Computing Facility (ALCF), the National Energy Research Scientific Computing Center (NERSC), and the National Center for Supercomputing Applications (NCSA).



Darshan eXtended Tracing (DXT) \cite{darshan_dxt} extends Darshan to provide fine-grain traces of the application I/O to study behaviors of a wide range of workloads in greater depth. The DXT module is disabled by default and can be switched on using an environment variable. The overhead introduced by this module is also minimal (around $1\%$ \cite
{darshan_dxt}). 

Recorder \cite{recorder} is a multi-level I/O tracing framework that captures I/O information from multiple levels of the I/O stack, such as HDF5, MPI-IO, and POSIX. It also requires no source code changes, and the user has control over which layers of the stack can be traced. Using library preloading, Recorder prioritizes itself over the standard functions using function interpositioning. Once Recorder is specified as a pre-loaded library, it intercepts HDF5, POSIX, and MPI-IO calls and reroutes these requests to the tracing routine. The tracing routine saves the function name, file name, entry timestamp, and the function parameters. All of this information is converted into a trace record and then compressed. The recent version of Recorder adds new features such as trace format optimization, visualizations, and improved tracing for POSIX \cite{9150354}. 

TAU \cite{TAU, Shende2011CharacterizingIP} is a portable and flexible integrated toolkit of performance instrumentation and analysis of HPC applications. Although TAU is not an I/O-specific profiling tool, it does provide different instrumentation techniques to analyze the I/O performance of the application. One of these approaches is pre-processor-based instrumentation, in which the header files are redefined to instrument POSIX I/O calls. It also performs MPI instrumentation using the PMPI interface.

Score-P \cite{10.1007/978-3-642-31476-6_7} is a performance measurement framework for profiling, tracing, and online analysis of HPC applications. Score-P provides different ways to instrument the application code, either through automatic compiler instrumentation or manual instrumentation. In the context of parallel I/O, Score-P can instrument POSIX I/O and MPI I/O routines. 

Mistral \cite{mistral} is a commercially available HPC I/O monitoring tool from Altair. Mistral, due to its lightweight, can be easily deployed on production systems and is flexible enough to monitor I/O patterns on large-scale HPC and cloud systems. It uses files called contracts to store the rules for monitoring and throttling I/O. These contracts can be modified at runtime by privileged users. With its system and storage agnostic features, Mistral can monitor the read, write, metadata use, and storage performance of each job and file system. Below is a list of data that Mistral collects: 

\begin{itemize}
\item Collect I/O data per job, user group, mount point, and host
\item Track read(), write(), and metadata operations
\item Track time spent in I/O for each application to get file system performance
\item Track statistics like the total, mean, and max time spent doing I/O per job
\item Tracks CPU and memory utilization 
\end{itemize}

Ellexus Breeze is another flexible and user-friendly offline analysis tool to help optimize and tune complex applications. It can also capture MPI-IO and POSIX calls and store the information related to them in binary trace files. Ellexus Breeze classifies I/O calls into three different categories: \textit{bad I/O} (e.g., small read and write, zero-byte I/O, backward seek), \textit{medium I/O} (e.g., large read and write, forward seek, delete operations), and \textit{good I/O} (e.g., medium-sized read and write, etc.).

\begin{table*}
  \begin{tabular}{l|ccc|ccc|c}
  \multicolumn{1}{c}{} & \multicolumn{3}{c}{\textbf{Type}} & \multicolumn{3}{c}{\textbf{Stack Layer}} & \multicolumn{1}{c}{} \\
     \textbf{Tool} & \rotatebox{90}{\textbf{\small Profiling}} & \rotatebox{90}{\textbf{\small Tracing}} & \rotatebox{90}{\textbf{\small Monitoring}} & \rotatebox{90}{\textbf{\small High Level}} & \rotatebox{90}{\textbf{\small Parallel Middleware}} & \rotatebox{90}{\textbf{\small Low Level}} &  \textbf{\small Visual Support} \\

    \toprule
     Darshan \cite{darshan} & \YES & \NO & \NO &  \YES & \YES & \YES & \YES\\
     IPM \cite{IPM} & \YES & \NO & \NO & \NO & \YES & \YES & -\\
      TAU \cite{TAU} & \YES  & \YES & \NO & \YES & \YES & \YES & \YES\\
     Score-P \cite{10.1007/978-3-642-31476-6_7} & \YES & \YES & \NO & \NO & \YES & \YES & \NO\\
     Darshan DXT \cite{darshan_dxt} & \NO & \YES & \NO & \NO & \YES & \YES &\NO\\
     Recorder \cite{recorder} & \NO & \YES & \NO & \YES & \YES & \YES &\YES\\
     RIOT I/O \cite{RIOT} & \NO & \YES & \NO & \NO & \YES & \YES & -\\
     ScalaIOTrace \cite{ScalaIOTrace} & \NO & \YES & \NO & \YES & \NO & \NO &\NO\\
     Mistral/Breeze \cite{mistral} & \NO & \NO & \YES & \NO & \YES & \YES &\YES\\
    \bottomrule
  \end{tabular}
  \caption{List of different application profiling, tracing, and monitoring tools and their characteristics}
\label{tab:app_profiling}
\end{table*}

RIOT I/O \cite{RIOT} is another I/O tracer that uses function interpositioning to intercept the POSIX and MPI-IO libraries storing timing and other information about each function call. Similarly, ScalaIOTrace \cite{ScalaIOTrace} is another tracing and analysis tool that collects trace data across multiple layers in the I/O stack. It also provides novel capabilities to automatically analyze and collect statistical information from the collected traces. IPM \cite{IPM}, developed at Lawrence Berkeley National Lab, is a portable profiling tool that profiles the performance aspects and resource utilization of parallel programs. By using an interposition layer, it captures all the calls between the application and the file system layer. 

\subsubsection{Holistic Characterization Tools}
Until now, this paper has looked at different I/O characterization and monitoring tools that instrument and monitor different layers in the I/O stack. These tools collect performance data for individual components in the I/O stack. However, just looking at this data individually does not convey the whole picture of the I/O performance of an application or system. In many cases, the data from different components and layers of the stack needs to be combined to get a more holistic view of the I/O behavior. This often requires involving an I/O expert in the loop to translate this disparate data, but not only is this practice labor-intensive, but it is also not sustainable as the system grows and scales. To tackle this issue, monitoring systems have been developed that provide a unified and detailed view of performance across the entire I/O stack by combining insights from the performance data collected by various component-level monitoring tools. 

The Total Knowledge of I/O (TOKIO) \cite{osti_1632125} is a framework that connects different component-level monitoring and characterization tools, combines their insights, and presents a single, holistic view of the I/O performance across the entire I/O stack, which can be further used by static analysis tools and user interfaces. The distinguishing factor in the TOKIO framework is that it provides a modular implementation that connects to whatever monitoring and profiling tools are available on the HPC system. It mainly collects data from the tools that profile application behavior, such as Darshan, and tools that look at storage system and network traffic and health. Connectors are the foundational layer of TOKIO. These are modular and independent components that provide an interface to connect with individual component-level tools, providing data from these tools. On top of connectors are TOKIO tools, which make this data collected by the connectors semantically closer to how analysis applications would like that data to be. These tools provide abstractions to hide the complexities of the underlying data source. High-level applications like command line tools, statistical analysis, and data analysis tools can easily connect with TOKIO interfaces to analyze the holistic data collected. 

Similar to TOKIO, the Unified Monitoring and Metrics interface (UMAMI) \cite{10.1145/3149393.3149395} also provides a normalized, holistic view of the I/O behavior of the application. With the component-level data already being collected by different tools at the application level and storage-system levels, this data is analyzed over one month and the changes in the metrics are shown in UMAMI. \citet{10.1145/2749246.2749269} presents a multiplatform study in which Darshan logs representing a combined total of six years of I/O behavior of a million jobs across three leading HPC systems are mined and analyzed. It studies the evolution of the I/O behavior of the application over time, and based on the findings, the study provides techniques to improve the I/O performance of an application.  

\begin{tcolorbox}[size=title, top=5px, bottom=5px, boxrule=0.1mm, arc=0.5mm, colback=white, colframe=black, fonttitle=\sffamily\bfseries, title=Summary \#3]
After generating the I/O workloads, the next step in parallel I/O evaluation and optimization is to profile and monitor the application’s I/O performance and use that data to understand the application’s I/O behavior and determine tuning strategies. A variety of tools are available for profiling, tracing, and monitoring I/O performance, and these tools are categorized into two groups: (1) Parallel File and Storage System Profiling Tools, and (2) Application Profiling Tools. These tools help analyze I/O behavior at different layers of the stack, from high-level parallel middleware to low-level I/O libraries like HDF5, MPI-IO, and POSIX. The section also highlights different holistic characterization and monitoring tools, such as TOKIO \cite{osti_1632125}, UMAMI \cite{10.1145/3149393.3149395}, and Beacon \cite{Beacon}. Each tool provides a different methodology to combine performance data from various profiling tools and present a complete view of the I/O performance of the application. The findings of this section are reported in Table \ref{tab:sys_profiling} and \ref{tab:app_profiling}.
\end{tcolorbox}

\section{Optimizing Parallel I/O}
\label{sec:parallel-io-optimization}
Until now, the paper has looked at various methodologies for parallel I/O evaluation on large-scale systems, which includes workload generation techniques, and profiling, tracing, and monitoring tools. When performance is slower than expected, end-users and system administrators rely on the profile and trace information these tools provide to identify the root causes of inefficiencies and bottlenecks \cite{ISC-paper}. Despite the availability of numerous tools that collect I/O metrics on production systems, identifying the specific I/O bottlenecks, understanding their root causes, and determining the appropriate optimizations to eliminate these bottlenecks remains a challenging task. Solving this challenge requires a thorough understanding of the different optimization techniques for the different layers of the I/O stack as well as knowledge of different I/O analysis and optimization tools. 

In this section, two key aspects of parallel I/O  optimization are discussed: (1) optimization strategies for parallel file and storage systems, as well as application performance, and (2) tools and research works that employ various analysis, modeling, and prediction techniques to assess and enhance I/O performance. Previous work \cite{previous_paper} has explored various analysis, modeling, and prediction techniques for optimizing parallel I/O. While \cite{previous_paper} provides valuable insights into parallel I/O optimization, this work surveys more papers and provides a comprehensive synthesis of the collected information. 

\subsection{Optimizing Parallel File and Storage System Performance}


\subsubsection{File Striping}

Lustre uses file striping to distribute the segments of a single file across several Object Storage Targets (OSTs). OSTs are a set of storage volumes that are stored on Object Storage Servers (OSS) which provide bulk storage for the Lustre. 
File striping is a critical optimization technique as it takes advantage of several OSSs and OSTs to increase the throughput of operations. File striping allows multiple clients to read/write different parts of the same file in parallel. Typically, large files are striped over multiple OSTs and striping small files can decrease performance. 

Three different tuning parameters control file striping: (1) \textit{stripe size}, (2) \textit{stripe count}, and (3) \textit{stripe offset}. Stripe size defines the size of the file stripes in bytes, stripe count defines the number of OSTs to stripe across, and strip offset defines the index of the OST where the first stripe of files will be written. The value assigned to these parameters can have varying effects on the I/O performance of the Lustre-based system. For example, \citet{inproceedings} explores how different combinations of these parameters can affect I/O performance and provides recommendations on what the optimal parameter configuration can be. The study determines that stripe size has little to no effect on the I/O performance of the system and sticking to the default stripe size is best. For
very small files (below 1 GB), the study observed modest improvements when using slightly larger stripe sizes (up to about 16 MB). The study also shows the effect of stripe count on I/O performance and determines that choosing a stripe count multiple of N (Number of Nodes) is the best strategy.

\paragraph{Request Scheduling}

On a large-scale supercomputer, there can be multiple HPC applications that communicate concurrently with the PFS to perform I/O operations. This concurrency can lead to performance issues and appropriate I/O scheduling techniques need to be applied across various levels of the I/O stack to mitigate this issue. There are a lot of I/O scheduling techniques applied at different levels of the stack (clients, I/O nodes, or data servers) that optimize concurrent access to the file system by organizing or reordering the requests. Let's assume there are two different applications that are issuing I/O requests to the file system. It is a possibility that when these requests arrive at the I/O forwarding layer or the file system, they are interleaved, hindering the performance. Even if multiple processes in one application are accessing a shared file contiguously, their requests to the file system might be perceived as non-contiguous because of interleaving. This phenomenon is called interference. 

There are a variety of I/O schedulers available and each has a different level of complexity. Some use simple algorithms such as First Come, First Serve (FIFO) \cite{Ohta2010OptimizationTA} while others like aIOLi \cite{article}, Object-Based Round Robin \cite{Qian2009A}, and Network Request Scheduler \cite{article3} are more complex. \citet{two-choice} presents a two-choice I/O scheduler that tracks the real-time performance of different storage servers to detect stragglers and uses that information to avoid placing I/O requests on stragglers. It builds upon the native two-choice algorithm and proposes collaborative probe and pre-assigns algorithms to improve performance. InterferenceRemoval \cite{10.1145/1810085.1810116} is another I/O scheduler that identifies those parts of the file that are involved in interfering accesses and replicates those portions to other data servers for better concurrency.

IOrchestrator \cite{5644884} is another I/O scheduler that exploits the spatiality of the application for request scheduling. It introduces the concept of reuse distance which is the time between two requests of the same program that have been sent to the data server. Assuming that these requests from the same program have a strong spatiality, IOrchestrator dedicates resources to this one program only. \citet{10.1145/2063384.2063407} uses a similar approach and introduces the concept of time windows. All the I/O requests forwarded to the storage server are regarded as a time series and are divided into fixed-size time windows. The algorithm sorts the I/O requests in a time window by application ID and assuming that each application has strong spatiality, the execution will be optimized as the requests belonging to the same application are executed at the same time. 

\citet{6546095} presents a hierarchical I/O scheduling technique for collective I/O. Collective I/O has been used widely to address the issues of a large number of I/O requests. As systems grow, the performance of collective I/O can also be significantly impacted as the shuffle cost increases because of highly concurrent data accesses. Hierarchical I/O scheduling (HIO)
algorithm addresses this problem by using shuffle cost analysis to schedule applications’ I/O requests for optimal performance. 

Cross-application interference can significantly impact the I/O performance of applications as there can be a lot of I/O requests with different sizes and requirements. In \citet{6877251}, a cross-application coordination technique to mitigate I/O interference is presented. To mitigate this issue, \citet{6877251} presents four different strategies: serializing, interrupting, interference, and dynamically adapting the best strategy based on the application's access pattern. \citet{5644883} proposes an adaptive algorithm to vary the I/O workload based on the performance of the file system. 

\subsection{Optimizing Application Performance}


\subsubsection{HDF5 Alignment}
HDF5 file alignment is a critical HDF5 optimization technique \cite{hdf5_tuning}. To understand HDF5 file alignment, it is crucial to understand how the application data is stored in datasets within HDF5 files. HDF5 datasets are composed of multi-dimensional arrays, where each array element can be of integer, floating-point, or a more complex data type. The multi-dimensional arrays in HDF5 datasets can have two types of composition: (1) contiguous or (2) chunked. 
In contiguous composition, all the dataset elements are stored contiguously within the file, whereas in chunked composition, all the dataset elements are stored in equal-sized chunks within the file, which enables fast compression and filtering operations as well as fast access to dataset elements. 

Many PFS perform optimally when data accesses fall on a particular chunk boundary \cite{hdf5_tuning}. The HDF5 library normally does not align data to a particular byte boundary, and instead packs data as tightly as possible. However, to match the requirements of the PFS and for optimal data accesses, the HDF5 library provides the H5Pset\_alignment API call which aligns all objects in a file over a particular size threshold. The H5Pset\_alignment API call involves two parameters: (1) \textit{threshold value}, and (2) \textit{alignment interval}. Any file object with size equal to or greater than the threshold bytes will be aligned on an address that is a multiple of the specified alignment. For optimal performance and mapping, each chunk's dimensions can be chosen so that when a parallel process accesses a subset of the dataset, it maps exactly to one chunk in the file. This optimization technique provided by the HDF5 library minimizes lock contention in the PFS and maximizes the file system throughput.

\subsubsection{HDF5 Metadata}
To store information about the datasets in the file, HDF5 files use metadata information \cite{hdf5_tuning}. This metadata is stored in the form of B-trees which maps dataset elements access requests for each chunk from array coordinates to file offsets. Maintaining the depth and breadth of the B-tree is important as an untuned B-tree can add a significant overhead. When the number of chunks in the dataset is known, H5Pset\_istore\_k API call can be used to tune the B-tree node width to exactly match the number of chunks. This API call reduces the number of I/O operations needed to bring the B-tree into memory ,which reduces the overall mapping cost from array coordinates to file offsets. 

\subsubsection{HDF5 Asynchronous I/O VOL connector}

Asynchronous I/O overlaps the I/O time with computation and communication, hence significantly reducing the I/O time of scientific applications \cite{https://doi.org/10.1002/cpe.8046}. The asynchronous I/O VOL connector was developed to enable asynchronous I/O for HDF5 operations. It works by maintaining two threads: (1) Application thread, and (2) Background thread. Argobots \cite{8082139}, a lightweight low-level threading framework, is used to manage these threads. It also maintains an async task queue to perform asynchronous I/O operations. The vol connector is compiled as a dynamically linked library (DLL) and then loaded through an environment variable by the HDF5 library. 

Asynchronous I/O can have two modes in HDF5: \textit{implicit} and \textit{explicit}. In implicit mode, minimal code changes are required, only requiring some environment variables to be set, but it has more limitations, such as synchronous read operations. In explicit mode, code changes are required such as replacing the HDF5 APIs with the EventSet APIs. This gives the application more control over the execution of the asynchronous operations and also provides a better mechanism for detecting errors \cite{https://doi.org/10.1002/cpe.8046}. 

\subsubsection{Request Aggregation and Reordering}

Request aggregation and reordering are two optimization techniques that are applied at the I/O middleware layer to transform the access patterns to be more suitable for the layers underneath. These optimization techniques include collective buffering and data sieving which are part of ROMIO \cite{romio}, which is a portable implementation of MPI-IO.

\citet{8752753} shows that small and random I/O requests harm the I/O performance of the system. For better performance, it is better to merge these small requests into larger fewer requests that span a large portion of the file. ROMIO uses data sieving \cite{750599} to make fewer requests to the file system. When a process makes small independent non-contiguous data requests, ROMIO does not access each data section separately. Let's assume that an MPI rank issues four non-contiguous data accesses. Instead of making four requests to the file server, using data sieving, ROMIO will read a single contiguous chunk of data starting from the first requested byte till the last request byte in a temporary buffer using a single call. Once we have the data in the buffer, ROMIO will extract the data needed by the process in the process's buffer.

One potential problem with data sieving is that there can be memory issues if the user requests a large amount of data to be read. This problem can further worsen if there are large holes in the data. To deal with this, ROMIO provides multiple user-controlled parameters to define the maximum amount of contiguous data that can be read into the buffer by the process. This includes parameters like ind\_rd\_buffer\_size and ind\_wr\_buffer\_size. If the data to be read is larger than the value defined by these parameters, data sieving is broken down into chunks, reading only as much data at a time defined by the parameter. 

The major advantage of data sieving is that it always reads the data in large chunks, therefore avoiding the cost of small file accesses but at the cost of reading more data. A potential caveat can be that the holes between the data are too large and can outweigh the cost of reading extra data to avoid small accesses. However, that is not the case most of the time. 

Two-phase I/O \cite{twophase} is another access pattern optimization technique originally proposed for distributed systems. In two-phase I/O, there are aggregator processes that perform collective read and write. Aggregators are processes that issue the I/O requests to the file system. In collective reads, the aggregator is responsible for the part of the file requested by the collective reads and then distributes the chunks of the file to the processes participating in collective reads. Similarly, in collective write, the aggregator gathers data from a subset of processes participating in collective write into contiguous chunks in memory and writes the aggregated data to the file system. ROMIO provides two user-defined tuning parameters to control collective I/O. These parameters are \textit{cb\_nodes} and \textit{cb\_buffer\_size}. cb\_nodes refers to the number of aggregators and cb\_buffer\_size refers to the maximum buffer size on each aggregator.

\begin{tcolorbox}[size=title, top=5px, bottom=5px, boxrule=0.1mm, arc=0.5mm, colback=white, colframe=black, fonttitle=\sffamily\bfseries, title=Summary \#4]
  This section introduces different optimization techniques for application and PFS performance. First, it discusses different tuning options for the Lustre file system such as changing the stripe settings. It also talks about various research scheduling algorithms and how they optimize the efficiency of the PFS. For application performance optimization, this section discusses some HDF5 and ROMIO optimizations as an example. These optimizations include HDF5 alignment and metadata settings, as well as different tuning parameters for ROMIO such as the collective buffer nodes and their size. 
\end{tcolorbox}

\subsection{Analyzing, Modeling, and Predicting I/O performance}

Once the profile and trace data are collected for the HPC I/O application using the different I/O profiling, tracing, and monitoring tools discussed in section \ref{sec:parallel-io-evaluation}, the next step in parallel I/O evaluation and optimization is to analyze this data to identify I/O issues, model I/O performance, and predict future I/O performance of the HPC system. This section looks at different tools and research work that have been done to analyze and optimize the I/O performance of an HPC system. Each tool and research work uses different analysis techniques to extract meaningful insights from the performance data. These techniques can be divided into three broad categories:  \textbf{Statistic and visual analysis}, \textbf{ML-Based analysis and prediction}, and \textbf{Trace replay-based analysis}. 

\begin{table*}
\centering
  \begin{tabular}{llp{2.3in}llll}
   \textbf{Tools/Research} & \textbf{Data/Benchmark} & \textbf{Analysis Goal} & \rotatebox{90} {\textbf{System Analysis}} & \rotatebox{90} {\textbf{Application Analysis}} & \rotatebox{90} {\textbf{Visual Support}} & \rotatebox{90}{\textbf{Feedback Support}}\\
    \toprule
     \citet{Yildiz2016OnTR} &  Grid'5000 \cite{1542730} & Root causes of I/O interference & \YES & \NO & \NO & \NO \\
  \citet{8665806} & Darshan, LMT, Slurm  & Short and long term performance variation & \YES & \YES & \NO & \NO\\
    IOMiner \cite{8514906} & Darshan, LMT, Slurm  & Sweep-line analysis function  & \YES & \YES & \NO & \NO\\
    \citet{10.1145/2749246.2749269} & Darshan & Insights on application I/O behavior & \NO & \YES & \NO & \NO \\
    \citet{7980043} & Custom trace data & Storage system performance analysis & \YES & \NO & \NO & \NO\\
    PyTokio \cite{osti_1632125} & Multiple sources & High-level abstractions for analysis & \YES & \YES & \NO & \NO \\
     GUIDE \cite{9926255} & Multiple sources & Storage system performance analysis & \YES & \NO & \NO & \NO \\
     DXT-Explorer \cite{dxt-explorer, ISC-paper} & DXT & Interactive visualization of DXT logs & \YES & \YES & \YES & \NO\\
     Drishti IO \cite{drishti, 10579210} & Darshan, Recorder & Performance diagnosis and recommendations & \YES & \YES & \NO & \YES\\
    \bottomrule
  \end{tabular}
  \caption{Different statistical analysis tools/research and their characteristics} 
\label{tab:stat_prediction}
\end{table*}

\subsubsection{Statistic and visual analysis}

The traditional way of analyzing I/O data is through applying statistics and analysis techniques to extract meaningful patterns from the I/O traces, I/O characterization profiles, and other logs. Different statistical techniques such as Arithmetic Mean, Standard Deviation, Linear Regression, Probabilities, etc. are applied to the data to classify, correlate, and extract meaningful patterns from it. Applying statistical analysis on HPC workloads and data requires in-depth knowledge of the HPC system and extensive human effort. 

\citet{Yildiz2016OnTR} presents an extensive experimental study exploring the root causes of I/O interference in HPC systems. It first identifies the different points of contention in an HPC storage system and evaluates how the application's access patterns, file system and network configuration, and storage devices affect interference. Based on the experiments carried out on the Grid’5000 testbed and the OrangeFS file system, the study highlights seven different root causes of I/O interference. \citet{8665806} investigates various I/O performance issues by studying and analyzing performance data collected over a year at two leadership high-performance computing centers. The study looks at transient and long-term trends in I/O performance variability by using different analysis techniques such as correlative analysis and financial market technical analysis techniques on time series I/O performance data to identify regions of interest. The insights provided in this paper can help broaden the scope of instrumentation and analysis tools. 

IOMiner \cite{8514906} is a large-scale analytics framework to analyze instrumentation data using a centralized storage schema that combines log data collected using different instrumentation tools and a sweep-line analysis function that identifies root causes of poor I/O performance of an application. The centralized storage schema is designed in this way it takes away any format difference that the different log data might have, making it query-friendly and easy to use for the sweep-line analysis function. One of the challenges that IOMiner addresses is how to mine useful information and insights from the performance data collected on supercomputers which can run as many as millions of jobs in a short time. To address this challenge, IOMiner uses a Python API for the Spark framework called PySpark which speeds up data analysis on large-scale systems using parallel processing. In the analysis phase, IOMiner provides an analysis function that looks at five different contributing factors to the application's poor I/O performance. These factors are:
\begin{itemize}
  \item Small I/O requests
  \item Nonconsecutive I/O requests
  \item Utilization of collective I/O
  \item Number of OSTs used by each job
  \item Contention level
\end{itemize}

\citet{10.1145/2749246.2749269} presents a comprehensive study of the I/O behavior of applications on three large-scale supercomputers. It analyzes the Darshan logs, spanning years and months, on the three different supercomputers. Through in-depth analysis of these logs, the paper looks at different aspects of the I/O performance of the applications on each supercomputer. For example, one of the analyses that the paper presents is a platform-wide analysis in which the performance of I/O workloads on the three supercomputers is studied and techniques are presented to identify underperforming apps. The analyses show that most of the apps on each supercomputer never exceed the platform peak I/O throughput and low I/O throughput is the norm. The paper also presents other insights such as discovering that a small number of jobs and apps mainly dominate each platform's I/O usage. 

\citet{7980043} presents a comprehensive study and analysis of the I/O performance of a production leadership-class storage system by collecting large amounts of trace data using the recorder \cite{recorder} tracing tool. After collecting the data, the study thoroughly analyzes this trace data to get insights into the performance and variability of the storage system and provides some feedback on how to resolve the issues encountered in the analyses. The paper looks at two main scenarios to understand I/O performance do the storage system: 1) \textit{Performance of I/O issued to a single OST}, and 2) \textit{Impact of concurrent access to single OST}. It also looks at other scenarios such as the effect of system caching on user-perceived performance and the effect of large-scale parallel I/O. Some of the key insights of the paper are:
\begin{itemize}
  \item Concurrent access to a single OST might not lead to performance degradation for specific write sizes
  \item To maintain high throughput, system caching is critical
  \item Interference can be generated along the I/O path because of large scale parallel I/O
  \item Imbalanced I/O traffic distribution among OSTs can lead to performance degradation 
\end{itemize}

PyTokio \cite{osti_1632125} is a Python implementation of the TOKIO framework discussed in previous sections. Pytokio makes holistic analysis of the I/O performance of parallel systems easier by providing connectors to interface with many commonly used monitoring tools. It also provides analysis routines to extract insights from the data collected from the different monitoring tools. One of the components of pytokio is tokio tools which provides a high-level abstraction for accessing the combined data collected from multiple monitoring tools, allowing different analysis tools to build upon pytokio. GUIDE (Grand Unified Information Directory Environment) \cite{9926255} is another framework that we discussed earlier which collects, federates, and analyzes log data from the Oak Ridge Leadership Computing Facility (OLCF). It also provides some analytics and visualizations of the storage system like looking at the Lustre OST usage over time, I/O block sizes and space efficiency, and I/O request size distribution. 

DXT Explorer \cite{dxt-explorer, ISC-paper} is one such dynamic web-based log analysis tool that visualizes Darshan DXT logs and helps understand the I/O behavior of applications. The tool automatically analyzes and parses Darshan DXT log data to generate different interactive visualizations focusing on different facets of the I/O performance of the application such as operations, transfer sizes, spatial locality, OST Usage, and I/O Phases. These interactive visualizations provide zoom in, zoom out capabilities which can aid researchers, developers, and end-users in investigating areas of interest, consequently helping in identifying the root causes of various I/O bottlenecks.

Drishti IO \cite{drishti, 10579210} is another command line analysis framework that analyzes Darshan logs to pinpoint the various root causes of the I/O problems, also providing a set of actionable recommendations to get rid of these bottlenecks and improve I/O performance. By using the counters collected in Darshan profiling logs, Drishti detects some common bottlenecks and classifies the I/O insights into four categories based on their impact. The tool also can pinpoint the exact line in the source code where changes need to be made to optimize performance. Based on this analysis, the tool generates a report, highlighting the issues in the application and providing a set of recommendations to solve those issues.

\subsubsection{ML-Based analysis and prediction}
\begin{table*}
\begin{threeparttable}[b]
\centering
  \begin{tabular}{lllll}
    \textbf{Tools/Research} & \textbf{Prediction Technique} & \textbf{Prediction Output} & \rotatebox{90}{\textbf{System Performance}} & \rotatebox{90}{\textbf{Application Performance}}\\
    \toprule
  Omnisc’IO \cite{Omnisc’IO} &  Grammer-based model  & Next I/O operation & \NO & \YES\\
\citet{8956059} &  Random forest approach & Execution time & \NO & \YES \\
\citet{Schmid_Kunkel_2016} & Artificial neural networks & File access times  & \YES & \NO\\
    \citet{8102186} & Six prediction models\tnote{1} & Storage system performance state & \YES & \NO\\
     IONET \cite{IONET} & Deep neural networks & Latency of each I/O & \YES & \NO\\
    \citet{7307576} & Gaussian process model & Collective I/O tuning & \NO & \YES\\
     \citet{osti_1311633} & Genetic algorithm & Best combination of parameters & \YES & \YES\\
     \citet{9150371} & Random forest & Collective I/O tuning & \NO & \YES\\
     \citet{article-reg} & Six prediction models\tnote{2} & I/O performance such as read throughput & \YES & \YES\\
    \bottomrule
  \end{tabular}
   \caption{Different ML-based analysis and prediction tools/research along with their characteristics} 
   \label{tab:ml_modeling}
        \begin{tablenotes}\small {
       \item [1] Classification and Regression Trees (CART), Naive Bayes (NB), Gradient Boosting (GBT), Support Vector Machines (SVM), Random Forests (RF), and Neural Networks (NN)
       \item [2] Linear, Polynomial, K-nearest neighbors, Gradient boosting random forest (GBDT), Random Forests (RF), Multilayers perceptron (MLP), and Convolutional neural network (CNN)
       }
     \end{tablenotes}
     \end{threeparttable}
\end{table*}

Predictive Analysis, as the name suggests, is the kind of analysis that predicts future events based on the current information provided. Such analysis deploys techniques like data mining, predictive modeling, and machine learning on the trace/log data to predict future performance. By building a predictive model using the trace data available, accurate predictions on the future performance of the HPC system can be made. Such an analysis also includes auto-tuning techniques to predict the best parameters for each layer of the HPC I/O stack for optimized performance. 

To predict future I/O operations, Omnisc’IO \cite{Omnisc’IO} presents a novel approach using a grammar-based model of the I/O behavior of the application. Not only does it predict when future I/O operations will occur, but it also tells where these operations will occur and how data will be accessed. The grammar-based model is built at runtime using an algorithm derived from Sequitur \cite{sequitur}. The way Omnisc’IO operates is that first, it gets the call stack of the program, associating each call with an integer called context symbols. Then it builds a grammar-based model from these context symbols. Once the grammar is built, it predicts future events by choosing a predictor from the grammar and associating each predictor with a weight.

\citet{8956059} presents a novel framework for modeling and predicting the execution times of MPI programs. The framework captures the syntax tree of the parallel program and automatically instruments it, collecting important features. The instrumentor, developed in clang, inserts detective code around loops, branches, assignments, and MPI communications, generating an instrumented version of the code. Once the features are collected, the goal is to find a correlation between the collected features and the execution time. This multivariate nonlinear regression problem is solved using a random forest approach. \citet{Schmid_Kunkel_2016} use machine learning with artificial neural networks to analyze and predict file access times of a Lustre file system. 

\citet{8102186} presents a lightweight parallel test harness to collect I/O data on HPC systems and monitor performance states on different storage subsystems. Then it formulates a machine learning model to predict the transitions between those performance states during runtime. Treating this as a classification problem, this work uses a classifier to predict which I/O state a future I/O operation is likely to encounter for an I/O path to an OSS. This work applies six commonly used machine learning classifiers to the dataset: classification and regression trees (CART), naive bayes (NB), gradient boosting (GBT), support vector machines (SVM), random forests (RF), and neural networks (NN). According to the results, SVM provides the best prediction accuracy. 

IONET \cite{IONET} is another ML-based I/O latency predictor that builds models of storage devices. It then uses this model to predict the latency of every I/O of a full workload running on a target storage cluster without actually running it on the cluster. It collects traces from various sources and industry partners and then designs an ML model to learn from these traces. The ML model uses various machine learning techniques such as Deep Neural Networks, Random Forest, Logistic Regression, etc. to predict I/O latency from these traces. \citet{7307576} presents a sensitivity-based modeling framework that predicts the I/O performance of the system and its variability as a function of application and file system characteristics using a Gaussian Process Model. 

\citet{osti_1311633} optimizes the I/O performance of applications using an autotuning framework. This autotuning framework can efficiently optimize the different layers of the I/O stack such as HDF5, MPI-IO, and Lustre without requiring any source code changes. The framework has two main components: \textit{H5evolve} and \textit{H5Tuner}. H5evolve uses a genetic algorithm to search the I/O parameter space to find the best parameter combinations and then uses H5Tuner to inject the new I/O parameters into the application with minimal user involvement. \citet{9150371} presents another machine learning-supported I/O autotuning framework to tune I/O parameters for different layers of the stack. 

\citet{article-reg} predicts I/O performance using a regression-based approach by integrating system logs from various sources. First, the framework builds a joint database to store logs from different sources and then selects the important features from these logs using different scoring and feature selection algorithms. Once the features are selected, six different regression algorithms are developed, which use these features to predict I/O performance.

\subsubsection{Trace replay-based analysis}

\begin{table*}
\centering
  \begin{tabular}{lp{1.8in}p{1.8in}ll}
    \textbf{Tools/Research} & \textbf{Technique} & \textbf{Goal} & \rotatebox{90}{\textbf{System Performance}} & \rotatebox{90}{\textbf{Application Performance}}\\
    \toprule
 \citet{HAO20191} & Trace compression and merging & Portable benchmark & \NO & \YES\\
 DwarfCode \cite{7098397} & Trace compression and merging & Portable benchmark & \NO & \YES\\
 \citet{10.1145/2832106.2832108} & Extrapolation & Single trace for an arbitrary ranks & \YES & \YES\\
  IOscope \cite{10.1007/978-3-030-02465-9_7} & Specific or ready-to-visualize traces & Decrease trace collection overhead & \YES & \NO\\
 SynchroTrace \cite{10.1145/3158642} & Architecture agnostic traces &  - & \NO & \YES \\
    \bottomrule
  \end{tabular}
   \caption{Different trace replay-based modeling tools/research and their characteristics} 
   \label{tab:trace_replay}
\end{table*}

Replay-based modeling is another form of modeling and analysis that relies on historical I/O traces or characterization data. These traces, which contain detailed information about the computation and I/O behavior of an HPC application, can be analyzed to replay the I/O behavior of the original application through I/O workload replication and benchmarks. These benchmarks and workloads can then be further used to predict the I/O performance of the original application. They can also be used to test how the original application will behave in different real-world deployments and scenarios. Apart from this, these replay-based models can be also used to generate workloads for storage systems. There are a variety of tools and research works that study replay-based modeling, some of which are discussed in this section.

\citet{HAO20191} presents a replay-based modeling framework to automatically generate portable benchmarks for I/O intensive parallel applications. It introduces a trace merging and trace compression algorithm which it uses to generate benchmarks of the original I/O application, mimicking the application's computation, communication, and I/O access patterns. These portable benchmarks can also be used to predict the I/O performance of the original application without the extra overhead as the benchmarks are a scaled-down version of the original I/O application. To compress the traces, the framework uses a suffix tree-based algorithm to reduce the redundant data introduced because of the loops. It extracts and compresses these loops to reduce the size of the trace by several orders of magnitude. After compressing the traces, they are merged and are used to generate the C code of the benchmark for the original I/O application. 

Another tool that uses a similar approach of trace recording, merging, compression, and portable benchmark generation is DwarfCode \cite{7098397}. This tool can accurately predict the performance of multicore systems including process-level coarse (MPI) application performance. Easily extendable to HPC or Cloud computing environments, the idea behind the tool is to provide platform-independent abstractions of real application behaviors using computation and communication traces. The way DwarfCode works is that it first uses a lightweight tracing engine to capture the computation and communication traces of an executing application. It analyzes those traces and finally generates a “dwarf code” automatically. The dwarf code is a portable benchmark of the real application that mimics the application's behavior. On platforms with a similar architecture, the running time of the portable benchmark is anticipated to scale proportionally with that of the original program. Finally, the dwarf code can be used to replay on a target platform to predict the application’s performance.

\citet{10.1145/2832106.2832108} presents an extrapolation and replay-based tool called ScalaIOExtrap and ScalaIOReplay. It builds upon the ScalaIOTrace \cite{ScalaIOTrace} tool and provides extended functionality for extrapolation and replay-based I/O analysis. It uses ScalaIOTrace to collect and analyze a small number of traces, determining the relationship between the different parameters and the number of ranks, and then uses ScalaIOExtrap to generate a single trace file for an arbitrary number of ranks. The experiments conducted with this tool show that the new I/O trace generated by ScalaIOExtrap retains the trace structure, I/O size, and the number of operations. It also preserves the event ordering and time accuracy in the new trace. This tool enables large-scale parallel I/O evaluation without actually executing the original application. 

IOscope \cite{10.1007/978-3-030-02465-9_7} is another tracing tool that solves the problem of high overhead incurred by collecting large traces across multiple layers of I/O stack by generating specific and ready-to-visualize traces of the applications' I/O patterns. SynchroTrace \cite{10.1145/3158642} is another trace-based simulation tool for multithreaded applications that generates dependency-aware architecture agnostic traces. It also provides a replay mechanism that is aware of these dependencies and can simulate synchronization actions to estimate the performance and power of chip multiprocessor systems (CMP). 

\begin{tcolorbox}[size=title, top=5px, bottom=5px, boxrule=0.1mm, arc=0.5mm, colback=white, colframe=black, fonttitle=\sffamily\bfseries, title=Summary \#5]
This section discusses the next step in evaluating and optimizing parallel I/O performance which is to analyze the performance data to identify issues, model I/O performance, and predict future performance. This process involves several techniques, including statistical analysis, machine learning-based analysis and prediction, and trace replay-based analysis. Statistical methods, such as mean, standard deviation, and regression, are traditionally used to extract patterns from I/O data, though this requires significant expertise and effort. Machine learning and predictive modeling techniques, on the other hand, help forecast future I/O performance and optimize system parameters for better performance. Trace replay-based analysis involves using historical I/O traces to replicate the application’s I/O behavior, enabling performance testing under different scenarios and predicting how the application will perform in various environments. These techniques, combined, offer comprehensive approaches for understanding and optimizing I/O performance in HPC systems.
\end{tcolorbox}

\section{Discussion and Gaps}
\label{sec:gaps}
This paper covers the full spectrum of parallel I/O evaluation, from workload generation and I/O profiling to tracing and optimization techniques. It presents a thorough overview of the different tools and methodologies that are available to the HPC community at each step of parallel I/O analysis and optimization. This section briefly highlights some of the current gaps and limitations in methods and tools for I/O workload generation, profiling, tracing, and optimization, identifying opportunities for improvement and further research in each area. 

In surveying various application and system benchmarks, as well as workload replication and simulation frameworks, a core limitation recognized in these tools and frameworks is their limited ability to accurately represent real-world workloads. While many of the benchmarks discussed in this paper can effectively simulate simple I/O access patterns, they struggle to mimic the I/O patterns and behaviors of complex workloads. Most of the existing benchmarks focus on mimicking specific access patterns and use cases, which does not align well with the complexity and variability of I/O behavior found in real-world workloads. Furthermore, existing benchmarks and proxy applications use static I/O patterns, which remain constant throughout the test, an approach that again is not representative of actual applications in which the I/O patterns can change dynamically throughout the execution of the application. Finally, the current state-of-the-art benchmarks, such as IOR \cite{IOR}, although useful for measuring performance in large-scale environments, do not scale effectively to match the demands of petascale and exascale systems.

Beyond gaps and limitations in the I/O workload generation techniques, there are also some limitations in the existing profiling, tracing, and monitoring tools. This survey reveals that aside from IOPro \cite{IOPro}, no other profiling or tracing tool provides end-to-end visibility across all layers of the I/O stack (application, middleware, file system, storage). Commonly used profiling and tracing tools such as Darshan \cite{darshan}, TAU \cite{TAU}, and Recorder \cite{recorder} have partial support for each layer of the I/O stack, often requiring a combination of multiple tools to capture data across the entire I/O stack. This poses a significant challenge, as combining data from multiple tools can lead to fragmented insights because of the differences in data format and granularity, making it difficult to develop a unified view of I/O performance. Additionally, most of the profiling, tracing, and monitoring tools provide information about the I/O behavior in the form of static snapshots or summaries, lacking analysis capabilities that suggest actionable guidance for optimizations. This limitation creates a gap between data collection and optimization, as users are left to manually interpret the metrics and implement performance optimizations, often relying on an I/O expert. Furthermore, while these tools are effective for post-mortem analysis, they lack support for real-time, adaptive optimization, which remains an open problem in this area.  

The paper also examines various optimization techniques for parallel file and storage systems, high-level libraries, and parallel libraries, with a specific focus on Lustre, HDF5, and MPI. A critical gap in the different optimization techniques suggested for these libraries is how changing the tuning parameters of one layer can affect the performance of any other layer, given the complex interdependencies between them. For example, how can changing the Lustre striping settings affect MPI-IO collective buffering performance, or how does changing in HDF5 dataset chunking influence PFS behavior? Addressing these challenges requires new methodologies and tools capable of capturing the cross-layer interactions within the I/O stack. While \citet{10579210} does address this challenge for the high-level I/O libraries and the parallel middleware, more research needs to be done to extend such solutions to address cross-layer dependencies across all levels of the I/O stack. Additionally, a notable limitation in the different research studies and tools presented in this paper, which focus on the analysis and optimization of I/O, is how the suggested optimization techniques translate to production systems. Most of the optimizations proposed by these studies are experimented with benchmarks or small-scale runs of the application, without evaluating their effectiveness in full-scale production environments and with real-life workloads. Addressing these limitations could bridge the gap between theoretical optimization strategies and practical, scalable solutions for large-scale HPC environments.

\section{Conclusion}
\label{sec:conclusion}
Understanding the I/O performance of large-scale workloads is increasingly challenging due to the ever-increasing complexities of the underlying parallel computing hardware and the advent of novel workloads comprising artificial intelligence, machine learning, and big data. In this context, one needs to have a systematic approach for evaluating and optimizing parallel I/O performance. This work attempts to provide that systematic approach by doing an extensive review of more than $130$ research papers, offering a comprehensive survey of the state-of-the-art parallel I/O evaluation and optimization tools and techniques for large-scale HPC systems. It thoroughly examines the HPC I/O stack, studying how data transformations reshape an application’s access pattern while I/O requests traverse the stack. Using a taxonomy to provide a structured overview of parallel I/O evaluation and optimization of large-scale HPC systems, this study presents a comprehensive list of workload generation techniques, profiling, and tracing tools for I/O characterization, all synthesized in a series of tables for easy reference. Finally, it explores various analysis and optimization methods-including statistical analysis, machine learning based analysis, and replay-based modeling, and discusses targeted optimization techniques for the different layers of the I/O stack.  

\section*{Acknowledgments}
This work was performed under the auspices of the U.S. Department of Energy by Lawrence Livermore National Laboratory under Contract DE-AC52-07NA27344. LLNL-JRNL-871650.

\bibliographystyle{ACM-Reference-Format}
\bibliography{bibliography}

\end{document}